\documentclass{ifacconf}
\usepackage{graphicx}      
\usepackage{natbib}        
\usepackage{textcomp}
\usepackage{xcolor}
\usepackage{mathtools}
\usepackage{comment}
\usepackage{url}
\usepackage{tikz}
\usepackage{adjustbox}
\usepackage{algorithm}
\usepackage{algpseudocode}
\usepackage{booktabs}
\usepackage{multirow}
\usepackage{siunitx}
\usepackage{subcaption}
\begin{document}
\begin{frontmatter}

\title{Iterative Youla-Kucera Loop Shaping For Precision Motion Control} 


\author[First]{Xiaohai Hu} 
\author[Second]{Jason Laks} 
\author[Second]{Guoxiao Guo}
\author[First]{Xu Chen}

\address[First]{Department of Mechanical Engineering, University of Washington, Seattle, WA 98195, USA. Email:{huxh, chx}@uw.edu}
\address[Second]{Western Digital Corporation, San Jose, CA, 95119, USA. {jason.laks, Guoxiao.Guo}@wdc.com}


\begin{abstract}                
This paper presents a numerically robust approach to multi-band disturbance rejection using an iterative Youla-Kucera parameterization technique. The proposed method offers precise control over shaping the frequency response of a feedback loop while maintaining numerical stability through a systematic design process. By implementing an iterative approach, we overcome a critical numerical issue in rejecting vibrations with multiple frequency bands. Meanwhile, our proposed modification of the all-stabilizing Youla-Kucera architecture enables intuitive design while respecting fundamental performance trade-offs and minimizing undesired waterbed amplifications. Numerical validation on a hard disk drive servo system demonstrates significant performance improvements, enabling enhanced positioning precision for increased storage density. The design methodology extends beyond storage systems to various high-precision control applications where multi-band disturbance rejection is critical.
\end{abstract}


\end{frontmatter}

\section{Introduction}
Precision control systems face significant challenges in simultaneously rejecting disturbances across multiple frequency bands while maintaining strong stability margins. Such a multi-band disturbance rejection problem is particularly critical in hard disk drives (HDDs), where positioning accuracy within 10 nanometers must be achieved despite various mechanical and aerodynamic disturbances \citep{yamaguchi2011high}.
Modern HDD servo architectures—ranging from dual-stage \citep{Pan2016TripleStageTS} to quadruple-stage designs \citep{atsumi2019qua}—must compensate for multiple disturbance sources that manifest as both repeatable runouts (RROs) at specific frequencies and non-repeatable runouts (NRROs) across broader frequency bands. As storage densities increase, positioning error tolerances become increasingly stringent, making effective multi-band disturbance rejection essential for performance.

Existing approaches to narrow-band disturbance rejection typically employ multiple second-order compensators targeted at specific frequencies. While these methods can effectively attenuate disturbances at their target frequencies, they often struggle with unintended interactions between frequency bands. The coupling between control actions at different frequencies can create unexpected peaks in the error sensitivity function at adjacent frequencies. Additionally, due to the fundamental waterbed effect described by Bode's integral theorem \citep{stein2003respect}, for most mechatronic systems, disturbance rejection at specific frequencies inevitably leads to error amplification elsewhere in the spectrum, requiring careful management of high-frequency noise amplification.
Although optimal control theory provides frameworks for minimum variance disturbance rejection given complete plant and disturbance models \citep{richard2012h2,gx2001optimal,chu2024optimal}, practical implementation requires addressing specific numerical challenges in shaping the sensitivity function. Loop shaping techniques offer direct control over frequency response characteristics \citep{wang2017,chen2015,chen2013acc} but have historically lacked systematic methods for managing multi-band interactions and numerical stability.
This paper presents a new approach based on Youla parameterization that offers three significant benefits. First, it automatically manages the phase and gain relationships in loop shaping, eliminating the need for manual adjustments that often plague traditional methods. Second, it provides better insight into how control actions at different frequency bands affect each other, allowing designers to anticipate and mitigate unwanted interactions between narrow-band control objectives. Finally, it gives engineers a clear visual representation of the waterbed effect at each design stage, making performance trade-offs more transparent and helping to guide the iterative design process.

Our methodology proceeds by formulating a sensitivity function loop shaping problem (Section \ref{sec:problem_formulation}), developing an iterative Youla Parameterization, and proposing a corresponding Q-filter design method (Section \ref{sec:youla_param}). We validate this approach through numerical case studies (Section \ref{sec:practicalimplemenation}) and demonstrate its practical application to an HDD (Section \ref{sec:casestudy}). While focused on HDDs, the proposed design process extends readily to precision manufacturing stages, optical systems (telescopes, laser steering), semiconductor lithography, and scientific instrumentation where simultaneous multi-band disturbance rejection is critical and traditional methods typically fail beyond 4-5 frequency bands.
\section{Problem Formulation}
\label{sec:problem_formulation}

We focus on loop shaping for single-input single-output (SISO) discrete-time systems. 
Let $\mathcal{S}$ denote the set of stable, proper rational transfer functions.
Given a discrete-time transfer function $P(z)$, we denote $P(e^{j\omega})$ as its frequency response, and $P(1) = P(e^{j\omega})|_{\omega = 0}$ as its DC gain.

Consider a plant $P(z)$ and a controller $C(z)$ in a negative feedback loop. We assume the loop gain $L(z) = P(z)C(z)$ is unity-gain stable—meaning that it maintains stability in unity gain feedback with sufficient phase margin at the unity gain frequency. This ensures that the resulting sensitivity function $S(z) = \frac{1}{1+L(z)}$ belongs to $\mathcal{S}$.

It is well understood from robust control and loop shaping, that the sensitivity function $S(z)$ is the transfer function from the reference input $r[k]$ to the error signal $e[k] (:= r[k]-y[k])$, while the complementary sensitivity function $T(z) = 1-S(z) = \frac{P(z)C(z)}{1+P(z)C(z)}$ connects the reference input to the system output $y[k]$. These functions serve as fundamental design parameters with the inherent constraint $S(z) + T(z) = 1$, which limits achievable performance. For effective disturbance rejection and robust performance, $S(e^{j\omega})$ should have small magnitudes at low frequencies, ideally with $S(1) = 0$ for perfect tracking of constant inputs. At high frequencies, where $L(e^{j\omega}) \approx 0$, we necessarily have $S(e^{j\omega}) \approx 1$.
More generally, for practical systems, performance enhancement in a feedback control design inevitably encounters fundamental limitations rooted from the waterbed effect and Bode's Integral Theorem, wherein sensitivity reduction in one frequency range necessitates increases elsewhere~\citep{mohtadi1990bode}. For our analysis, we assume the loop transfer function satisfies the relative degree condition: $\partial(N) \leq \partial(D) - 2$, where $N(z)$ and $D(z)$ are
the numerator and 
denominator polynomials of the transfer function, respectively.
This condition is satisfied by general motion control systems with motor current/voltage as input and position as output, and ensures appropriate rolloff characteristics of the control system at high frequencies.

The proposed iterative loop shaping forms a series of control components $C_k(z)$'s for rejecting a large amount of narrow-band disturbances, while 
allowing designers to manage the performance-stability tradeoff for each $C_k(z)$. 
Algorithmically, this is achieved by leveraging Youla-Kucera (YK) parameterization, aka Youla parameterization, and a novel design of the series of add-on controllers from a baseline feedback loop. 
In particular, we transition from a plant \(P\)-focused design to one centered on the loop gain \(L\), with careful attention to delay terms in discrete-time control and the preservation of stability conditions during controller synthesis. From there, the design concept can be applied to general dynamic systems.


\section{Relevant Background on Inversion-Based YK Parameterization}\label{sec:youla_param}

%

\textbf{Loop Shaping with Standard YK Parameterization:}
Given an SISO plant \(P(z)\) and a coprime factorization of \(P(z) = N(z)/D(z)\), e.g., \(N(z)=z^{-1}\) and \(D(z)=1-z^{-1}\) for \(P(z)=\frac{z^{-1}}{1-z^{-1}}\), YK parameterization enables a unified expression of all the possible feedback controllers \(C_{all}\) that stabilize \(P(z)\) (see, e.g., ~\citep{vidyasagar2011control}). In particular, for our case with a stabilizing baseline controller \(C(z)\), if \(C(z)\) is stable itself, then \(C_{all}\) can be expressed as:
\begin{equation}
\label{eq:stdYK}
C_{all}(z) = \frac{C(z)+D(z)Q(z)}{1-N(z)Q(z)}
\end{equation}
where we simply require \(Q\in\mathcal{S}\) and \(1-N(\infty)Q(\infty)=0\). The latter condition is often rather easy to satisfy (consider the earlier example where \(N(z)=z^{-1}\) and hence \(N(\infty)=0\). For more examples, see ~\citep{chen2013adaptive}).

The YK parameterization reduces the search for all the stabilizing controllers (which can be any rational transfer functions) to a single parameter \(Q(z)\), which is only needed to be a stable rational transfer function (i.e., in the set \(\mathcal{S}\)). Furthermore, the new sensitivity function becomes 
\begin{equation}
\tilde{S}(z) = \frac{1}{1 + P(z)C_{all}(z)} = \frac{1 - N(z)Q(z)}{1 + P(z)C(z)}
\label{eq:sen_Q}
\end{equation}
By designing $Q(z)$ such that $|1 - N(z)Q(z)| \approx 0$ over frequency bands of interest, we can significantly reduce the sensitivity and hence improve disturbance rejection in those frequency bands. 

\begin{figure}[htb!]
    \centering
\includegraphics[width=0.9\linewidth]{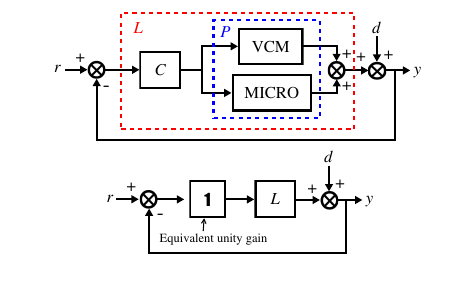}
    \caption{Equivalent Block Diagrams of Dual-Actuator Single-Output Control Systems}
    \label{fig:equivalentloop}
\end{figure}

\textbf{Loop Shaping with Inversion-Based YK Parameterization:}
The inversion-based YK parameterization offers significant advantages in sensitivity function shaping, particularly for disturbance rejection tasks. In particular, since a baseline controller is already designed in many practical control systems, if we focus on the loop transfer function \(L(z)\) instead of \(P\), we can have an equivalent feedback loop with \(L(z)\) as the fictitious plant and the unity gain \(1\) as the feedback controller~\citep{XuChen_MISOYK}.
This is highly advantageous for dual- or multi-input single-output systems such as the HDDs, as the intricate baseline control system can now be simplified as a single quantity as shown in the red dashed line in Fig. \ref{fig:equivalentloop}.

In this reduced-complexity YK framework, the inverse-based controller parameterization frames the transfer functions as:
\begin{equation}
L(z) \approx \frac{z^{-m}}{z^{-m} \hat{L}^{-1}(z)}, \quad C(z) = 1
\label{eq:inverse_yk_approx}
\end{equation}
where \(\hat L(z)^{-1}\) is the stable nominal inverse model of the loop transfer function and \(m\) is the relative degree of \(\hat L(z)\).
Applying Eq. (\ref{eq:stdYK}) with \(N(z)=z^{-m}\), \(D(z)=z^{-m} \hat{L}^{-1}(z)\), and \(C(z)=1\) leads to the Q parameterization of all the stabilizing controllers:
\begin{equation}
{C}_{Q}(z) = \frac{1 + z^{-m} \hat{L}^{-1}(z) Q(z)}{1 - z^{-m} Q(z)}
\label{eq:q_param_controller}
\end{equation}

Here, $\hat{L}^{-1}(z)$ is often implemented using stable-inversion techniques~\citep{chen2013adaptive} such as the Zero Phase Error Tracking Controller (ZPETC) ~\citep{zpetc1987}) and \(H_\infty\) optimal inversion~\citep{chen_optInverse2019}. The parameter $m$ ensures causality of \(D(z)\) by making \(z^{-m}\hat L^{-1}(z)\) to be realizable.

The key advantage of this formulation is the simplified sensitivity function:
\begin{align}
\tilde{S}(z) &= \frac{1}{1 + L(z)C_{Q}(z)} 
\approx \frac{1-z^{-m}Q(z)}{1+L(z)} \nonumber \\
&= S_0(z)(1-z^{-m}Q(z))
\label{eq:inverse_yk_sensitivity}
\end{align}
where \(S_0(z):=1/(1+L(z))\).
This form decouples the Q-filter design from the plant dynamics (except for causality considerations), offering more intuitive and direct shaping of the closed-loop response. More specifically, the sensitivity function becomes a product of the original sensitivity $S_0(z)$ and a shaping term $1-z^{-m}Q(z)$, allowing us to focus our design efforts on crafting appropriate Q-filters to achieve desired rejection at specific frequencies.

By selecting $Q(z)$ such that the frequency response $z^{-m}Q(z)|_{z=e^{j\omega}} \approx 1$ at frequencies of interest, we can create notches in the sensitivity function, effectively rejecting disturbances at those frequencies. The depth and width of these notches can be precisely controlled through proper Q-filter design, which we will explore in subsequent sections.
\section{Proposed Iterative YK Loop Shaping}
While the YK-parameterization based loop shaping is intuitive and powerful, when the controller order is high, the design of \(Q(z)\) must be numerically carefully designed. Instead of implementing a high-order \(Q\) directly (see an example in Section \ref{sec:numericalCond}), the proposed iterative YK loop shaping gradually builds up the loop shape for rejecting multiple band-limited disturbances.
More specifically, given a stable baseline feedback loop with \(L(z)\) and the unity feedback gain \(1\), we add the first desired loop shaping via \(\tilde C_1(z)\) by
\begin{equation}
\tilde{C}_{1}(z) = \frac{1 + z^{-m} \hat{L}^{-1}(z) Q_1(z)}{1 - z^{-m} Q_1(z)}
\end{equation}
and subsequently add more controls via designing 
	\begin{equation}\label{eq:ykinverse}
		\tilde C_k (z)= \frac{\tilde C_{k-1}(z) + z^{-m}\hat L^{-1}(z)Q_k(z)}{1 - z^{-m}Q_k(z)}, \quad Q_k \in \mathcal{S}.
	\end{equation}
We will elaborate the detailed design rationale soon, after 
discussing the general design framework of the Q filter in the next subsection, 


\label{sec:practicalimplemenation}

\subsection{Q Filter Design Framework}
For the inverse YK scheme, when $r(k) = 0$ and $z^{-m}Q(z)|_{z=e^{j\omega_i}} \approx 1$, the output of the Q filter approximates the input disturbance at $\omega_i$, effectively neutralizing the disturbance and forming a disturbance observer~\citep{chen2013adaptive}. This time-domain interpretation provides valuable insights for practical implementation. Our primary objectives are twofold: achieving near-zero sensitivity at specified frequencies for effective disturbance rejection while maintaining the original sensitivity function $S_0$ at other frequencies, thereby minimizing the waterbed effect at higher frequencies.
Let $\omega_i, i= 1,\dots,n$ represent a set of distinct frequencies in radians. We define the disturbance characteristic polynomial $A_\zeta(z)$ as:
\begin{equation}
\label{eq:Azeta}
\begin{aligned}
A_\zeta(z) &= \prod_{i=1}^n (1 - 2\zeta \cos(\omega_i) z^{-1} + \zeta^2 z^{-2})\\
&=\prod_{i=1}^n (1 - \zeta e^{j\omega_i} z^{-1})(1 - \zeta e^{-j\omega_i} z^{-1})
\end{aligned}
\end{equation}
where $\zeta \in \{\alpha, \beta\}$, with $0 < \alpha < \beta \leq 1$, and $\beta$ approaching or equal to 1. 
Noting that the roots of \(A_\zeta(z)\) are \(\zeta e^{\pm j\omega_i}\), \(i=1,2,\dots\),
we let 
\begin{equation}
    \label{eq:yk}
1-z^{-m} Q(z)=\frac{A_\beta(z)}{A_\alpha(z)} K(z)
\end{equation}
where \(K(z) = k_0+k_1 z^{-1}+k_2 z^{-2}+...\) depends on the value of \(m\) and is solved by matching coefficients on both sides of the equation (e.g., \(K(z)=1\) if \(m=1\)).
Then, because of the locations of the roots in \(A_\beta(z)\) and \(A_\alpha(z)\), at each $\omega_i$, a notch is created in the magnitude response of the sensitivity function.  When $\beta=1$, the frequency response of \(A_\beta(z)\) in Eq. (\ref{eq:Azeta}) and hence Eq. (\ref{eq:yk}) are exactly 0 at \(\omega_i\)'s, namely, perfect disturbance rejection is achieved at $\omega_i$ with $Q\left(e^{j \omega_i}\right)=e^{j m \omega_i}$.
 
\begin{figure}[htb]
    \centering
    \includegraphics[scale=0.5, trim={100pt 195pt 120pt 190pt}, clip]{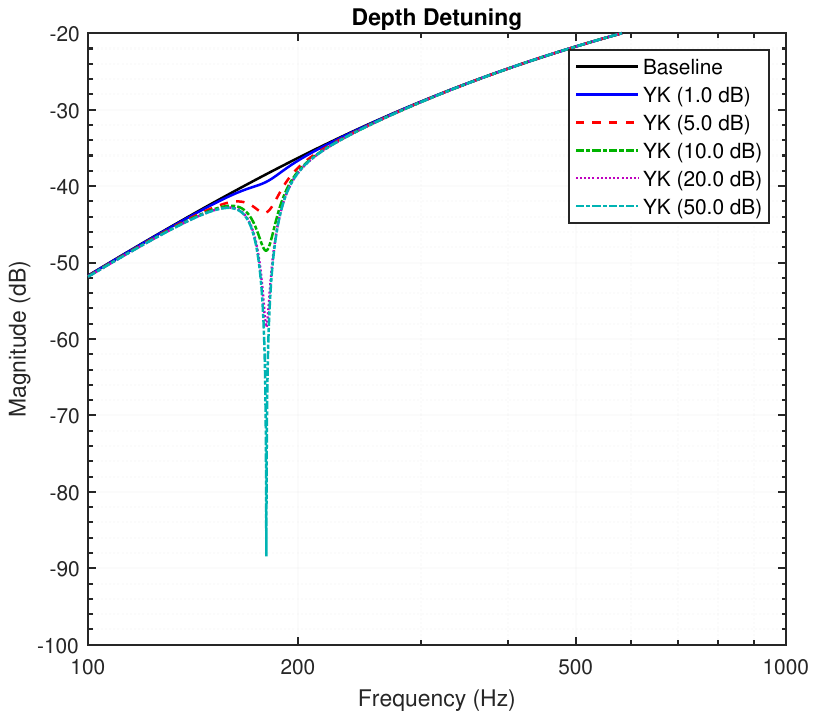}
    \caption{Sensitivity function tuning with attenuation factor g for 180 Hz notch (30 Hz bandwidth). Color progression from black (g=0, baseline) to dark blue ($g\approx1$) shows 1-50 dB attenuation levels while preserving response outside target band.}
    \label{fig:depth}
\end{figure}

For additional freedom in fine-tuning the loop shape, we introduce two principal mechanisms:\\
\textbf{Depth tuning:}
Here, we introduce a scaled Youla parameter $\tilde{Q}(z) = g \cdot Q(z)$, where $0 \leq g \leq 1$. This scaling reduces $\left|Q\left(e^{j \omega}\right)\right|$, causing $\left|1 - e^{-j m \omega} \tilde{Q}\left(e^{j \omega}\right)\right|$ to approach unity outside the pass bands, effectively mitigating waterbed amplification.
Fig. \ref{fig:depth} demonstrates this relationship through a single-notch design centered at 180 Hz with a 30 Hz bandwidth, illustrating the sensitivity function's response across various values of $g$.



\textbf{Width Adjustment:}
For the bandpass design, the design of IIR notch filters, as elaborated in~\citep{IIR1991}, allows for independent tuning of the notch frequency and attenuation bandwidth. The relationship between the attenuation filter denominator coefficient $\alpha$ and the 3-dB attenuation bandwidth $B$ is: 
\begin{equation}
    \sin \alpha = \frac{1-\tan (B / 2)}{1+\tan (B / 2)}
\end{equation}
This formulation enables precise control over both the notch frequency and attenuation bandwidth, facilitating the design of narrow-band rejection filters with independently adjustable characteristics.


\subsection{Numerical Conditioning in Multi-band Disturbance Rejection Control}
\label{sec:numericalCond}

\begin{figure}[t]
  \centering
  \begin{tabular}{@{}c@{\hspace{2mm}}c@{}}
    \includegraphics[width=0.5\columnwidth, trim={100 210 130 200}, clip]{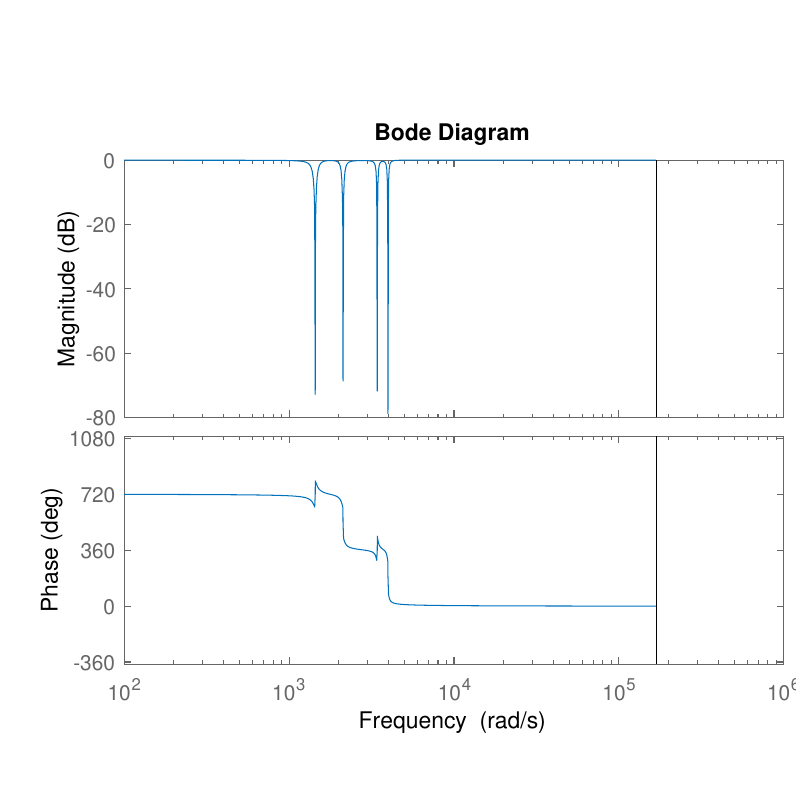} &
    \includegraphics[width=0.5\columnwidth, trim={100 210 130 200}, clip]{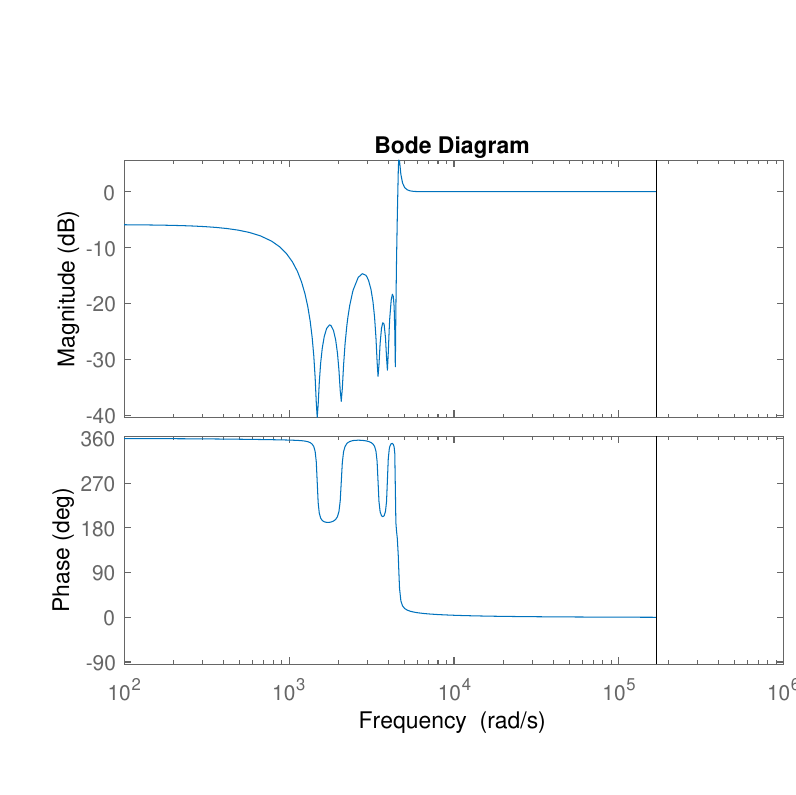} \\
    \footnotesize{(a) Four-Band Design} & \footnotesize{(b) Five-Band Design}
  \end{tabular}
\caption{Numerical conditioning comparison: (a) Stable four-band design with well-defined notches, (b) Five-band design showing numerical instability and erratic response, demonstrating traditional method limitations beyond 4-5 bands.}
  \label{fig:sensitivity-comparison}
\end{figure}

In the synthesis of cascaded second-order sections for multi-band disturbance rejection, numerical precision becomes paramount, particularly when pole-zero pairs approach the unit circle. Consider a design  of Eq. (\ref{eq:yk}) with a nominal bandwidth B = 20 Hz and target disturbance frequencies at $\{229, 338, 545, 633, 740\}$ Hz. 
The theoretical design positions poles at $\alpha e^{\pm j\omega_i}$ ($\alpha = \arcsin{\frac{1-\tan (B/2)}{1+\tan (B/2)}}$) with magnitude $\|\cdot\| = 0.9988$ (where $\|\cdot\|$ denotes the modulus of complex numbers). When implementing with 7-digit precision, significant pole migration occurs, with certain poles exceeding unity magnitude ($\|\cdot\| > 1$), violating discrete-time stability criteria. While 15-digit precision demonstrates improved pole location maintenance, it still exhibits numerical artifacts, in configurations exceeding four frequency bands.

This numerical degeneracy manifests in the frequency response of the added sensitivity $1 - z^{-m}Q(z)$, as shown in Fig.~\ref{fig:sensitivity-comparison}. The four-band implementation (Fig.~\ref{fig:sensitivity-comparison} a) maintains numerical stability, whereas the five-band case (Fig.~\ref{fig:sensitivity-comparison} b) generates poles and zeros outside the unit circle, despite both designs sharing identical $\alpha$ values and bandwidth specifications. These stability violations persist irrespective of pole placement $\alpha e^{\pm j\omega_i}$, indicating fundamental numerical conditioning issues rather than design limitations.

\subsection{Iterative Design of Youla Kucera Parameterization}
\begin{figure}[htb!]
    \centering
    \includegraphics[width=0.9\linewidth]{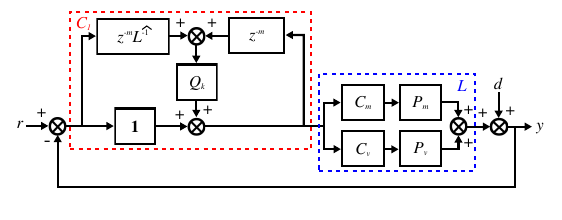}
    \caption{Block diagram for first iteration of YK design showing initial unity gain stage}
    \label{fig:first_iter}
\end{figure}

\begin{figure}[htb!]
    \centering
    \includegraphics[width=0.9\linewidth]{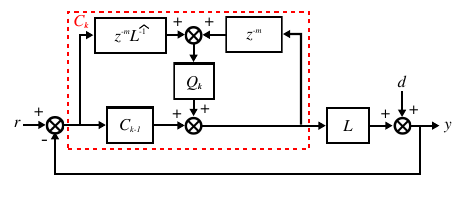}
    \caption{Method 1: Iterative controller modification approach for multi-stage disturbance rejection implementation}
    \label{fig:iter_controller}
\end{figure}

We propose an iterative YK parameterization methodology to address the limitations of existing approaches. Figs.~\ref{fig:first_iter} and \ref{fig:iter_controller} illustrate this new framework, wherein the blue dashed box represents the initial unity gain stage. Subsequent iterations involve the incremental addition of 1-2 notches in the loop shape, with each pass refining the system through this designated area. A pseudocode is shown in Table \ref{alg:pesudo_yk_iter}

The proposed implementation enables iterative computation of the Q filter for multiple frequency groups, updating the controller and sensitivity function at each step. The reduced-order controller from each iteration is accumulated to form the final controller capable of rejecting disturbances at all specified frequencies.
\begin{algorithm}
\caption{Iterative Q Design with Order Reduction}
\label{alg:pesudo_yk_iter}
\begin{algorithmic}[1]
\Require $n_g$ (\# of target frequencies), $r_{\text{red}}$ (reduction order), $\omega_{\text{groups}}$, $L$, $m$, $\omega_b$  (attenuation bandwidth)
\Ensure Reduced controllers and sensitivity functions
\State $\mathcal{S}_{\text{results}} \leftarrow [\mathcal{S}_{\text{baseline}}]$; $L_{\text{inverse}} \gets \textsc{ZPETC}(L)$ \\ \Comment{ZPETC: Zero Phase Error Tracking Control}
\State $C_{\text{total}} \leftarrow 1$; $L_{\text{current}} \leftarrow L$
\For{$i = 1$ \textbf{to} $n_g$}
    \State $C_{\text{all\_previous}} \leftarrow 1$; $g_{\text{factor}} \gets \textsc{GetGroupFactor}(i)$ \\ \Comment{Determines attenuation factor for group $i$}
    \For{each $\omega_j \in \omega_{\text{groups}}$}
        \State $Q_{\text{YK}} \gets \textsc{DesignQFilter}(\omega_j, \omega_b)$ $\times$ $g_{\text{factor}}$ 
        \State $C_{\text{new}} \gets \frac{C_{\text{all\_previous}} + z^{-m}L^{-1}_{\text{inverse}}Q_{\text{YK}}}{1 - z^{-m}Q_{\text{YK}}}$ 
        \State $C_{\text{reduced}} \gets \textsc{ReduceOrder}(C_{\text{new}}, r_{\text{red}})$ \\ \Comment{Model reduction to order $r_{\text{red}}$}
        \State $C_{\text{all\_previous}} \gets C_{\text{all\_previous}} \times C_{\text{reduced}}$
        \State $L_{\text{current}} \gets L_{\text{current}} \times C_{\text{reduced}}$
        \State $\mathcal{S}_{\text{new}} \gets \frac{1}{1 + L_{\text{current}}} \times (1 - z^{-m}Q_{\text{YK}})$
    \EndFor
    \State $C_{\text{total}} \gets C_{\text{all\_previous}}$; $\mathcal{S}_{\text{results}}.\textsc{append}(\mathcal{S}_{\text{new}})$
\EndFor
\end{algorithmic}
\end{algorithm}

\begin{figure}[htb!]
    \centering
    \includegraphics[width=0.9\linewidth]{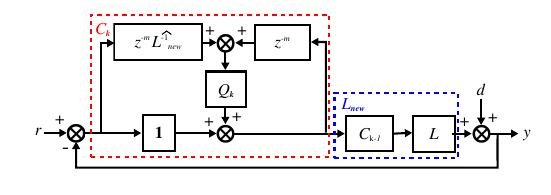}
    \caption{Method 2: Cascaded loop gain shaping alternative for multi-stage implementation}
    \label{fig:cascaded_loop}
\end{figure}
An alternative method, depicted in Fig. \ref{fig:cascaded_loop}, incorporates the controller into the loop gain at each iteration, treating them collectively as a new loop. However, this approach requires repeated inversion of the plant, leading to potential issues such as the accumulation of estimation errors introduced by the inversion approximation and a dramatic increase in the order of the plant as the process progresses.
\begin{figure}[htb]
    \centering
    \includegraphics[scale=0.4, trim={100pt 195pt 0pt 175pt}, clip]{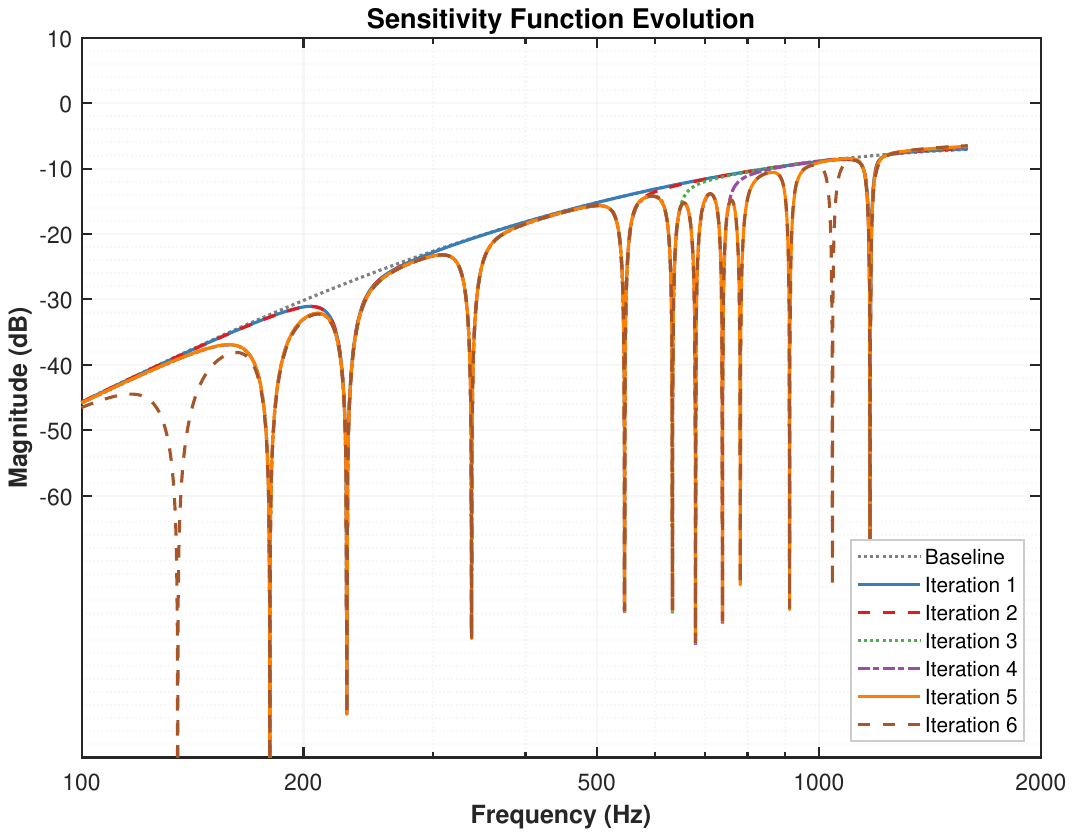}
    \caption{Sensitivity evolution across six iterations, with each iteration adding two notches for a total of twelve targeted frequency bands.}
    \label{fig:S_evolution}
    
\end{figure}
As demonstrated in Fig. \ref{fig:S_evolution}, here we use 6 stages of iterations and each time we add 2 notches, resulting in a total of 12 notches to attenuation at the desired frequency range. 

The proposed approach offers two notable advantages:
As illustrated in Fig. \ref{fig:iter_controller}, the iterative process eliminates the need for repeated factorization of previous loops, with only the antecedent controller compensation being considered in subsequent iterations.

This approach effectively transforms the complex multi-band control problem into a more manageable Q filter design task. The resulting framework offers both theoretical rigor and practical implementability, making it well-suited for robust control system applications.

\section{Case Study}
\label{sec:casestudy}
Consider a dual-stage HDD structure where a voice coil motor (VCM) and a micro-actuator are combined for enhanced positioning. A simplified block diagram is shown in Fig.~\ref{fig:equivalentloop}. Let $P = [P_1, P_2]$ represent the DISO plant, and $C = [C_1, C_2]^T$ denote the single-input dual-output (SIDO) baseline controller.
Then
\(    L = P_1C_1 + P_2C_2
\).
A comparative analysis of sensitivity functions elucidates the relative performance of our proposed method in relation to the established ESPRC approach, where second-order controllers are combined to provide sensitivity compensation, similar to the method described by~\citep{bashash2018tradeoffs}.

\begin{figure}[htb]
    \centering
    \includegraphics[scale=0.55, trim={120pt 195pt 120pt 195pt}, clip]{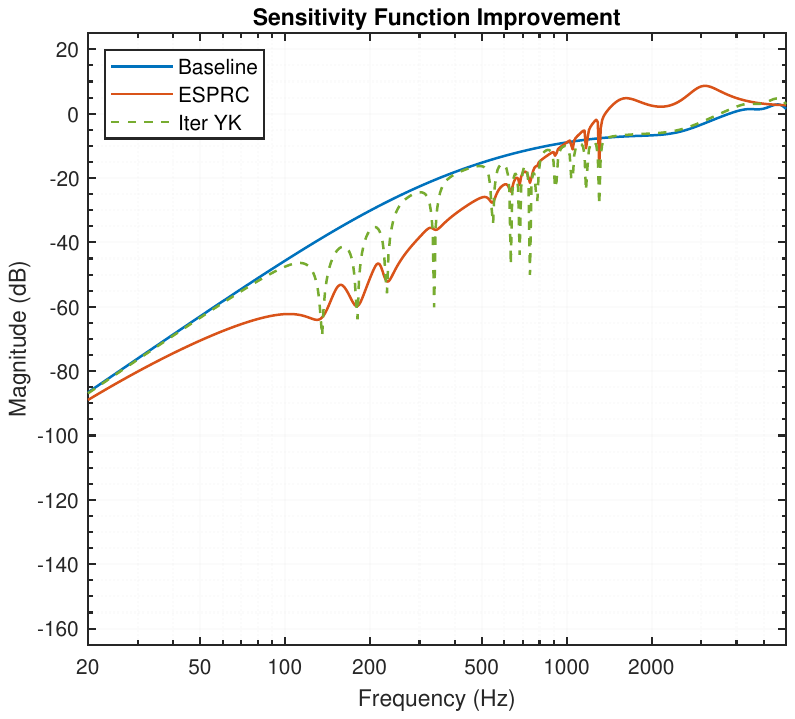}
    \caption{Loop shaping results comparing iterative Youla-Kucera parameterization method against ESPRC approach for multi-frequency vibration control in dual-stage HDD positioning system}
    \label{fig:yk_compare}
\end{figure}
\begin{figure}[htb]
    \centering
    \includegraphics[scale=0.55, trim={120pt 195pt 120pt 195pt}, clip]{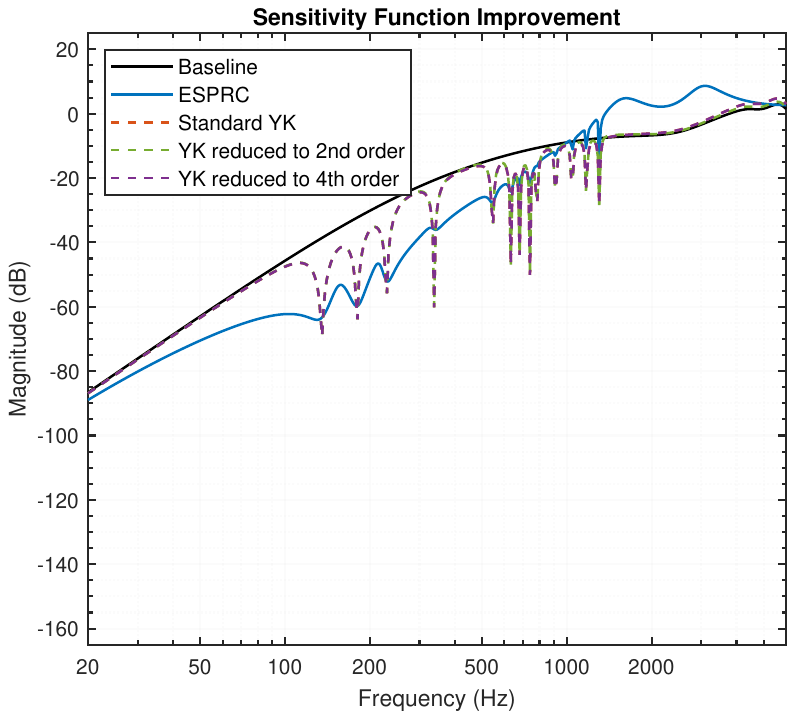}
    \caption{Controller order reduction comparison showing performance consistency between full-order (240th), 4th-order reduced (48th), and 2nd-order reduced (24th) controllers}
    \label{fig:red_order}
\end{figure}

The experimental results presented in Fig.~\ref{fig:yk_compare} demonstrate that our iterative YK parameterization method achieves superior attenuation performance compared to the ESPRC approach regarding deep attenuation of a large amount of band-limited vibrations. 
In addition, a distinguishing feature of the proposed approach is its effective management of the waterbed effect at higher frequencies. This capability is particularly significant in multi-frequency vibration control applications.

To enhance implementation feasibility and computational efficiency, we implemented a systematic order reduction strategy at each iteration step. This reduction methodology was executed at two distinct levels: a conservative 4th-order reduction and a more aggressive 2nd-order reduction, resulting in final controllers of 48th and 24th order, respectively.

As illustrated in Fig.~\ref{fig:red_order}, the reduced-order controllers maintain performance characteristics remarkably consistent with their full-order counterpart. This is particularly evident in the frequency response plots, where the full-order controller (orange trace) and the 4th-order reduced version (purple trace) demonstrate nearly identical behavioral patterns. 



\section{Conclusion}
This paper has presented an iterative approach to multi-band disturbance rejection using Youla-Kucera parameterization that addresses traditional numerical limitations in high-order control design. Our methodology successfully implements up to twelve notch filters while maintaining system stability. The experimental validation on a dual-stage HDD servo system confirmed superior attenuation performance across all target frequency bands while effectively managing high-frequency amplification. 
This framework establishes a foundation for addressing complex disturbance rejection problems in precision motion control applications where traditional methods face fundamental limitations. The methodology extends directly to precision manufacturing, optical systems, semiconductor fabrication, and scientific instrumentation requiring simultaneous rejection of multiple narrow-band disturbances. Future work could explore the integration of this method with optimization algorithms to further optimize the design process by user-specified optimization cost function. The modeling part of HDDs (\cite{hu2024state}) and identifying the disturbance frequencies is also important.

\bibliography{ifacconf}             
\end{document}